\documentclass[runningheads]{llncs}
\usepackage{amssymb, adjustbox, multirow, booktabs}
\usepackage{graphicx}
\usepackage{makecell,caption}
\usepackage[belowskip=-15pt,aboveskip=0pt]{caption}
\usepackage{tabularray}
\usepackage{capt-of}
\usepackage{hyperref, footnote}
\begin{document}
\title{SonoSAMTrack - Segment and Track Anything on Ultrasound Images}
%
%
%
\author{Hariharan Ravishankar,
Rohan Patil,
Vikram Melapudi,
Harsh Suthar,
Stephan Anzengruber,
Parminder Bhatia,
Kass-Hout Taha,
Pavan Annangi
}	

\authorrunning{Hariharan Ravishankar et al.}
\institute{GE HealthCare}
%
%
\maketitle              
\vspace{-0.2in}
\begin{abstract}
In this paper, we present SonoSAMTrack - that combines a promptable foundational model for segmenting  objects of interest on ultrasound images called SonoSAM
\cite{SonoSAM}, with a state-of-the art contour tracking model to propogate segmentations on 2D+t and 3D ultrasound datasets. Fine-tuned and tested exclusively on a rich, diverse set of objects from $\approx200$k ultrasound image-mask pairs, SonoSAM demonstrates state-of-the-art performance on $7$ unseen ultrasound data-sets, outperforming competing methods by a significant margin. We also extend SonoSAM to 2-D +t applications and demonstrate superior performance making it a valuable tool for generating dense annotations and segmentation of anatomical structures in clinical workflows.  Further, to increase practical utility of the work, we propose a two-step process of fine-tuning followed by knowledge distillation to a smaller footprint model without comprising the performance. We present detailed qualitative and quantitative comparisons of SonoSAM with state-of-the-art methods showcasing efficacy of the method. This is followed by demonstrating the reduction in number of clicks in a dense video annotation problem of adult cardiac ultrasound chamber segmentation using SonoSAMTrack.

\vspace{-0.1in}
\end{abstract}

\begin{figure}[!t]
\centering
\includegraphics[width=0.9\textwidth]{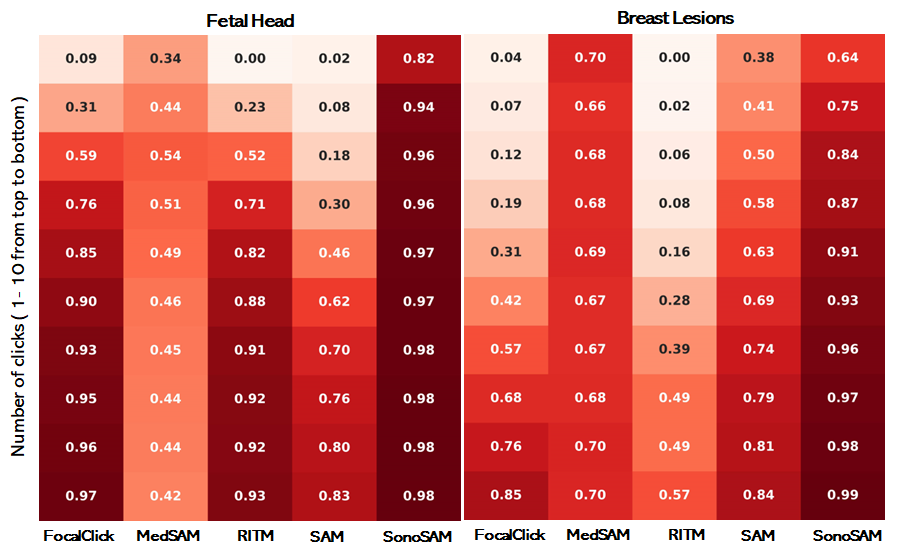}
\includegraphics[width=0.9\textwidth]{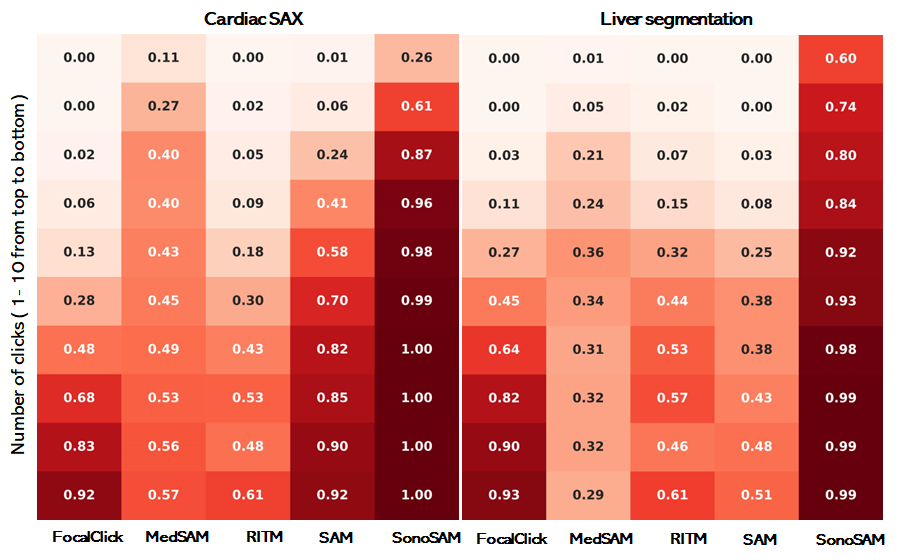}
\caption{Figure shows distribution of number of clicks versus percentage of images on which average DICE overlap exceeds 80$\%$, on $4$ different anatomies.} \label{fig:fig1}
\end{figure}


\begin{figure*}[!t]
\centering
\includegraphics[width = 1\textwidth]{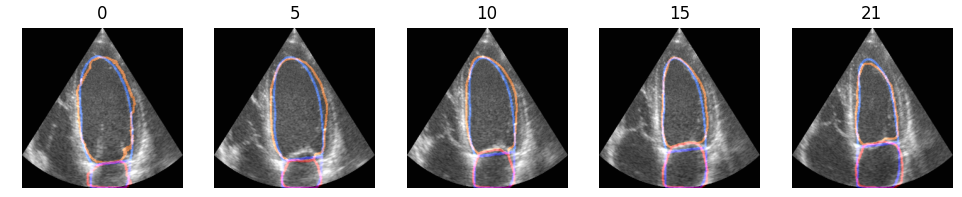}
\caption {Tracking results on a dataset with SonoSAM and deAoT \cite{track} on a 4-chamber ultrasound cine-loop on 5 frames showing Left Ventricle and Left Atrium segmentation. Legend: 1) Ground-truth : blue 2) Prediction : orange  }
\label{fig:track1}
\end{figure*}

\vspace{-0.25in}
\section{Introduction}
\vspace{-0.1in}
In many of the AI-powered ultrasound imaging applications, semantic segmentation of objects is of fundamental importance. 
While, popular DL architectures like U-Net\cite{unet} achieve state-of-the-art (SOTA) performance, the biggest bottleneck is in getting annotated data. Obtaining exact contour markings of objects of interest in ultrasound, mandates involvement of experts with clinical knowledge, is often tedious and time-consuming. The issue is exacerbated in 3-D or 2-D+t volumes, where getting dense contour marking for multiple objects across all the frames for a subject alone is extremely challenging. 
Recently, AI-powered tools have become popular for assisting object annotation in natural images. By learning on large number of image-mask pairs in the order of billions \cite{SAM}, these models learn the concept of  ``objectness" and function as generic, class-agnostic object segmentors. Models like FocalClick \cite{FocalClick}, Segment anything (SAM) \cite{SAM} have advanced ``promptable" segmentation, where user supplies prompts  and  models will automatically delineate objects of interest. The different types of user prompts are scribbles \cite{scribbles1,scribbles2}, bounding box \cite{bbox1,bbox2}, extreme points \cite{extreme}, clicks (most explored owing to ease of use), texts and images \cite{CLIP}. 
Among these options, the clicks have been most explored due to their simplicity and ease of use. Bounding box and extreme points demand more effort from users to capture the extents accurately; scribbles pose difficulty in simulating user behavior during training. Text and image based prompts require additional annotations. Early click based segmentation approaches formulated the task as an optimization problem of obtaining similar, connected region with reference to the region identified by the user prompt. These early approaches utilized graph based approaches \cite{GraphCut}, gaussian mixture models \cite{GrabCut}, and other related methods. Broadly these methods rely on strongly specified representation of semantic information and thus would demand large amount of user prompts to complete the task. One of the earlier attempts in click-based segmentation framework was presented in \cite{DeepCut}.The method utilized input image and  user clicks represented as distance maps as additional input for a CNN to predict a segmentation mask \cite{DeepCut}.


Major breakthrough in click-based segmentation was presented in RITM \cite{RITM}, using iterative sampling to generate clicks during training \cite{iterativeClick}. FocalClick \cite{FocalClick} utilized localized inference strategy to further improve accuracy. SimpleClick \cite{SimpleClick} explored the use of vision transformers for interactive segmentation. Segment anything model (SAM) \cite{SAM} extended this to multiple prompts, and trained on massive data (1 billion images) to enable several mainstream applications including zero-shot segmentation. SAM has emerged as the gold standard for promptable segmentation owing to its success in multiple domains \cite{SAM}.
Major breakthrough in click-based segmentation was presented in RITM method \cite{RITM} which significantly improved the effectiveness of learning based approaches with the use of an iterative sampling strategy to generate clicks during training \cite{iterativeClick}and use of high quality images, annotations. FocalClick \cite{FocalClick} utilized localized inference strategy to further improve accuracy, responsiveness and lowered compute requirements. SimpleClick \cite{SimpleClick} explored the use of vision transformers for interactive segmentation. Segment anything model (SAM) \cite{SAM} extended this to multiple prompts, via encodings, and trained on massive data (1 billion images) to enable several mainstream applications including zero-shot segmentation. SAM has emerged as the gold standard for promptable segmentation owing to its success in multiple domains \cite{SAM}.

Utility of SAM\cite{SAM} in medical imaging has been assessed recently in \cite{mazurowski2023segment,medsam,medperspectives}.
In an extensive experimental study \cite{mazurowski2023segment}, the authors find that while SAM obtains reasonable performance on different modalities, the performance is the poorest on ultrasound. This behavior is to be expected since ultrasound possesses unique characteristics like presence of scan cone, poor image quality and unique texture with speckles. Fig. \ref{fig:fig1}a shows inefficiency of SAM and other methods on different anatomies, where after each click, our method outperforms consistently all the other methods. 

An early attempt at developing a foundational model for medical imaging was proposed in MedSAM \cite{medsam}, where authors finetuned SAM on image-mask pairs obtained from 11 modalities including ultrasound, using only bounding box prompts. While MedSAM outperforms SAM on first prompt, it is still extremely inadequate, fails consistently with more clicks, and is often poorer than SAM on ultrasound data. This has prompted the community to speculate that either modality-specific or organ-specific models \cite{medperspectives} are the optimal granularity for foundation models to be of practical utility in the clinical community.

With this motivation, we present SonoSAM \cite{SonoSAM}: An ultrasound modality-specific segment anything foundation model trained with a set of $\approx200$k ultrasound image-mask pairs and tested on $\approx35$k ultrasound images. Specifically in Ultrasound imaging, the data can come as 2D, 2D+t or 3D acquisitions based on the probe and clinical applications. It is of utmost value to have a solution that can work on the different acquisition modes to enable practical utility of the work. To this effect, we present SonoSAMTrack that combines SonoSAM with state-of-the art tracking algorithms, to extend our model to 2D+t ultrasound datasets.

The key aspects of this work are as follows:
\vspace{-0.05in}
\begin{itemize}
	\item a first of its kind foundational model exclusively for ultrasound images enabling promptable segmentation,
	\item extensions to 2-D + t ( 3D ) use-cases with SOTA tracking methods,
	\item demonstration of state-of-the-art performance on $7$ unseen ultrasound datasets,
	\item demonstration of reduction in time-saving with combined segmentation and tracking model on adult cardiac ultrasound datasets,
	\item development of a deployable low footprint model with knowledge distillation.
	
\end{itemize}
\vspace{-0.1in}
\section{Technical Details}
\vspace{-0.05in}
\subsection{Architecture}
\vspace{-0.05in}
For semantic segmentation, there are essentially two types of backbones in practice - 1) hierarchical backbone: predominantly CNN-based \cite{unet} which learn coefficients that exploit local image content, use downsampling layers and aggregation to capture global information. 2) Plain backbone: boosted by the success of vision transformers (ViT) in other problems, segmentation architectures without pyramidal feature aggregation architecture have become popular. In SimpleClick \cite{simple}, authors proposed architecture with plain backbone of various sizes - ViT\_b, ViT\_l, ViT\_h which correspond to base, large and huge vision transformers with 90M, 300M and 600M parameters respectively. These architectures were further used in building SAM \cite{SAM} models. Owing to the success of these  backbones, we fine-tuned SonoSAM models starting from ViT\_b model of SAM \cite{SAM}.

\vspace{-0.1in}
\subsection{Fine-tuning strategies}
\begin{figure}[!t]
\centering
\label{stages}
\includegraphics[width=\textwidth]{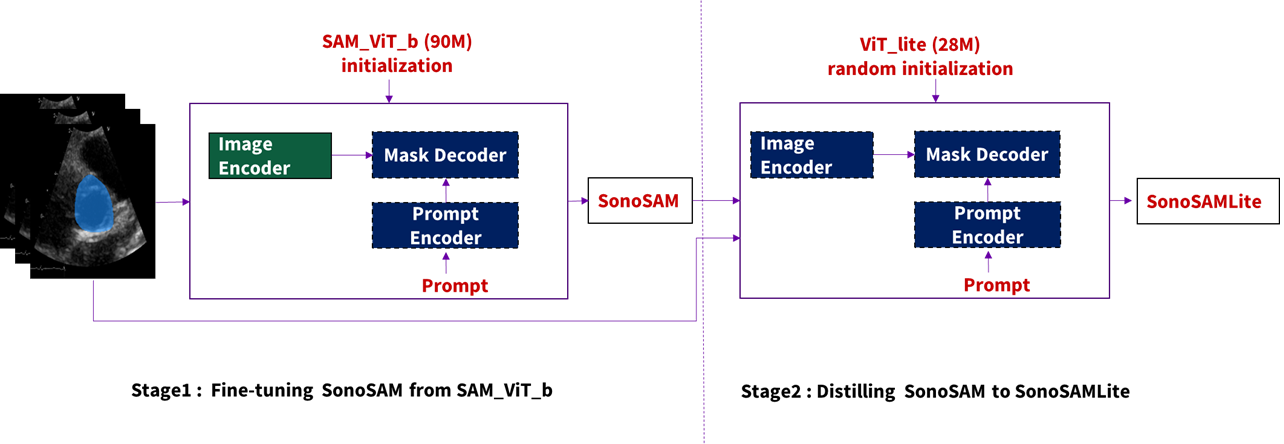}
\caption{Two stage process of building SonoSAM and SonoSAMLite. - Green boxes depict frozen weights and blue boxes capture fine-tuned weights.} \label{finetuningarchs}
\end{figure}

\subsubsection{a) Partial Fine-tuning - Domain specific decoder:}
SAM models are made up of $3$ sub-blocks namely Image encoder, prompt encoder and mask decoder. In terms of number of parameters, image encoder is heaviest containing almost $>90\%$ of parameter count. In the first stage (Fig. 2), we freeze the image encoder and fine-tune the prompt encoder and mask decoder on ultrasound images. Since, SAM has been trained on billions of natural images, we hypothesize that the image encoder of SAM should have reasonable generalization capabilities when it comes to encoding ultrasound images. However, since the concept of ``objectness" changes drastically in ultrasound data, we proceed to fine-tune the mask decoder - leading to ultrasound specific decoder which learns to produce contours of objects in ultrasound. We refer to this model as SonoSAM. We demonstrate that domain-specific decoder solution, while being practical from learning perspective, is also extremely adequate from performance perspective. SonoSAM is trained on ultrasound image-mask pairs, with DiceFocal loss proposed in \cite{DFLoss}, which is linear combination of focal loss and dice loss. We utilize Adam optimizer with initial learning rate of 1e-4 and decay of 0.5 with every 25 epochs.
\vspace{-0.12in}

\subsubsection{b) Knowledge Distillation:}
\begin{figure}[!t]
\centering
\includegraphics[width=\textwidth]{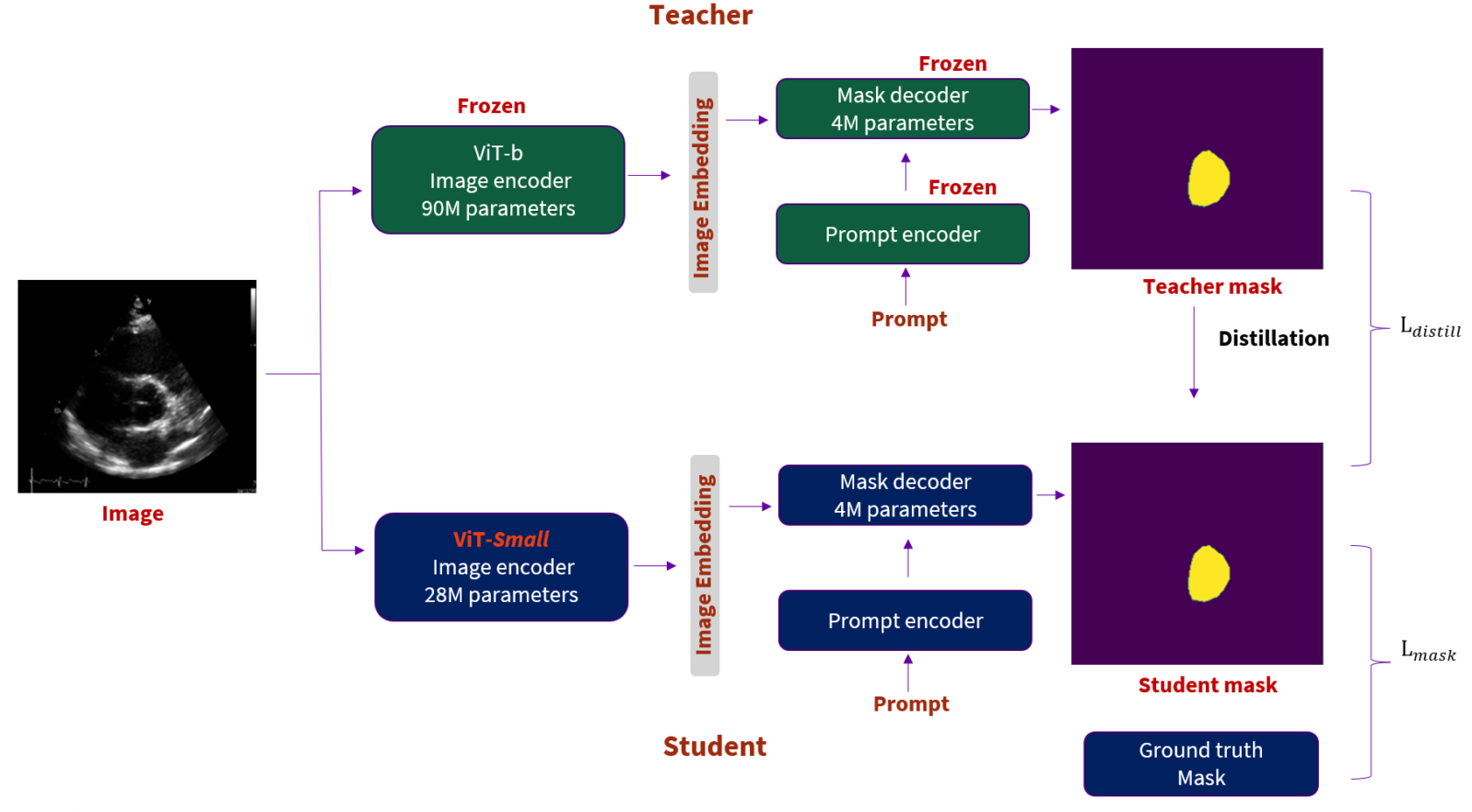}
\caption{Figure shows training of teacher-student model for SonoSAMLite. - Green boxes depict frozen weights and blue boxes capture fine-tuned weights.} 
\label{fig:finetuningSonoSAMLite}
\end{figure}

To enable practical utility of SonoSAM as a general purpose segmentation model for Ultrasound applications, it is desirable to have a model of a reasonable size that can be realistically deployed on devices or scanners.  However, the existing foundation models trained on billions of images consume significant amount of memory and compute.  While building a foundation model for a targeted healthcare domain of Ultrasound, it is desirable to have a model of a reasonable size that can be realistically deployed on devices or scanners, and can be used as a general purpose segmentation model for Ultrasound applications.  However, the existing foundation models trained on billions of images consume significant amount of memory and compute. 
This calls for a model that can demonstrate reasonably good and generalized segmentation performance as compared to SonoSAM, albeit with considerably lesser number of parameters. that allows it to be small enough to be deployed to scanners or medical edge devices. Further, one of the important practical limitations of foundation models is the requirement of a significant amount of memory and compute resources. With the availability of around 1.1B annotated images in case of SAM \cite{SAM}, one can argue that a sufficiently large number of parameters are necessary to be able to learn on a dataset of this scale. However, while building a foundation model for a targeted healthcare domain of Ultrasound, it is desirable to have a model of a reasonable size so that it can be realistically deployed on devices or scanners, and can be used as a general purpose segmentation model for Ultrasound applications. This calls for a model that can demonstrate reasonably good and generalized segmentation performance as compared to SonoSAM, albeit with considerably lesser number of parameters that allows it to be small enough to be deployed to scanners or medical edge devices.
To address this, in stage 2 , as shown in Figure ~\ref{fig:finetuningSonoSAMLite}, we use knowledge distillation \cite{distill}  to build a relatively light-weight student model SonoSAMLite, with SonoSAM as the corresponding teacher. The image encoder of SonoSAMLite is a lighter ViT-based architecture, resulting in a model which is one-third the size of the smallest ViT-b variant. We leverage the same architecture for the other two components - prompt encoder and mask decoder.  For learning, we use a weighted combination of mask loss between student and ground truth (eqn. 1), and a distribution-based distillation loss between student and teacher predictions. We use weighted loss made up of DiceFocal loss \cite{DFLoss} as $L_{mask}$, and KL-Divergence loss $L_{distill}$ for our experiments, with $\alpha=0.1$. 
\vspace{-0.05in}
\begin{equation}
	L_{student} = (1-\alpha) * L_{mask}(\hat{y}, y) + \alpha * L_{distill}(\hat{y}, y_{SonoSAM})
	\label{eqn:distill}
\end{equation}
Similar to the teacher, the student model is also trained with an iterative training strategy (Section. 2.3) where prompts are iteratively sampled based on the error regions of the model predictions. We employ a best-mask distillation strategy, where the output of the teacher corresponding to the iteration reporting the best mask is used for distilling the student. We hypothesize that this allows the student model to have consistent distillation targets and ground-truth targets across iterations and enables stable learning for the student. For training SonoSAM$_{student}$, we follow a similar iterative training strategy, where we simulate clicks or bounding box prompts sampled according to the predicted masks across iterations.

\vspace{0.2in}


\subsection{SonoSAMTrack workflow}

\begin{figure*}
\centering
\includegraphics[width = 1.0\textwidth]{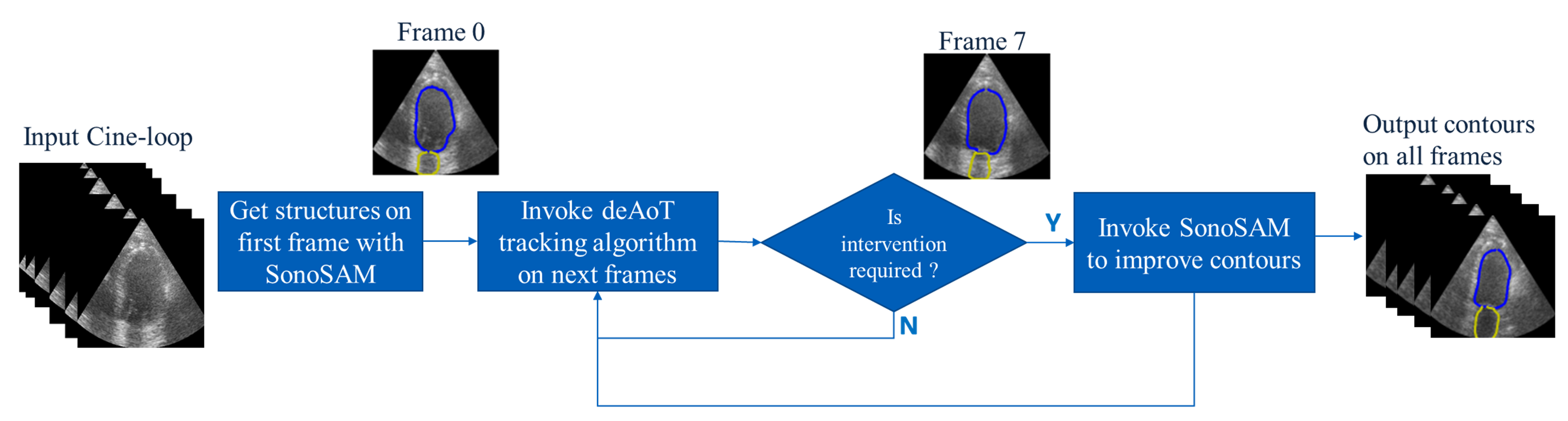}
\caption{Tracking framework for obtaining loop-level multi-structure segmentations using combination of SonoSAM and deAoT \cite{track} tracking algorithm.}
\label{fig:track}
\end{figure*}

Despite its practical significance, application of medical foundational models to 3-D has not receive much attention. In this section, we present initial results of our evaluations of SonoSAM for automated loop-level object segmentation. To address this gap without having to explicitly train a tracking network on case by case basis, we propose SonoSAMTrack that combines SonoSAM with a SOTA tracking algorithm.  To propogate segmentation contours over frames, we utilize a tracking algorithm presented in \cite{track} to track the objects. The deAoT method presented in \cite{track} proposes an effective method of hierarchical propagation for video object segmentation. From independent experiments, we found this method to work well on ultrasound images. The methodology of the SonoSAMTrack workflow is shown in Figure~\ref{fig:track} and works as follows: On the first frame, we invoke SonoSAM and generate satisfactory contours for all objects using user provided click prompts. On subsequent frames,  We closely monitor the DSC overlap with ground-truth and invoke SonoSAM only when DSC falls below 90\% - similar to how expert will intervene when contours have to adjusted. The proposed human in the loop system combining SonoSAM with SOTA tracking model can significantly reduce time taken to perform dense segmentations on 2D+t and 3D ultrasound datasets.

\vspace{-0.1in}

\subsection{Pre-processing}
The existing SAM models work on 3-channel RGB images which is suitable for natural images. For normalization of images, these models utilize pre-computed channel-wise mean and standard deviation obtained from ImageNet database. Firstly, ultrasound images are single channel grayscale images with intensity distribution distinctly different from natural images. To compensate for these factors, we compute statistics on ultrasound images from training data and use them for normalization, followed by replicating the normalized image to three channels. To accommodate images of different sizes, we initially resize the images to standard size. Further, we employ standard image augmentation techniques to increase data diversity during training.

\subsection{Iterative training mimicking user-interaction}
\vspace{-0.05in}
Inspired by the success of training mechanisms presented in \cite{RITM,FocalClick,SAM}, we follow an iterative prompt selection strategy to simulate  human interaction during training of both SonoSAM and SonoSAMLite. On every image, a foreground point or a bounding box is chosen with equal probability is chosen as the first prompt. To simulate a practical environment, where the user may not provide the exact prompt, we add controlled noise to both point and bounding box prompts. For point prompt, a pixel with a controlled jitter around the foreground centroid is selected at random. Similarly, for a bounding box prompt, the box co-ordinates are altered on either side to make a loosely fitting bounding box. For both type of prompts, the maximum jitter added is around 20\%. For the subsequent iterations, point prompts are uniformly sampled based on the error map between predicted mask and ground truth. Depending upon the dominant type of mistakes - false positive or false negative, a new negative or positive point is chosen in error region appropriately. 
%
\begin{table*}[]
\renewcommand{\arraystretch}{1.5}
    \resizebox{0.48\columnwidth}{!}{
    \begin{tabular}{|c|c|c|c|c|c|c|}		    
		\multicolumn{2}{l}{} & \multicolumn{2}{l}{\textbf{Training Data-set}} & \multicolumn{2}{l}{}\\		
		\toprule
    \textbf{Num} & \textbf{Anatomy} & \textbf{\#Images} & \textbf{Objects} & \textbf{Probe} & \textbf{Device} \\    
		\midrule
    1     & Fetal Heart 4ch  & 2500 & 5 & 2D & Voluson E8/E10  \\
		\midrule
    2     & Fetal Heart 3VT  & 2120 & 4 & 2D & Voluson E8/E10   \\
		\midrule
    3     & Fetal Thorax  & 8300 & 1 & 2D & Voluson E8/E10     \\
		\midrule
    4     & Gynaecology  & 87000 & 3 & 3D & Voluson E10     \\
		\midrule
    5     & Kidney  & 1500 & 1 & 2D & Logiq E10      \\
		\midrule
    6     & Liver  & 3700 & 1 & 2D & Logiq E10     \\
		\midrule
    7     & Bile Duct & 150 & 1 & 2D & Logiq E10   \\
		\midrule
    8     & Female Pelvic & 682 & 1 & 4D & Voluson E10     \\
		\midrule
    9     & Thyroid  & 450 & 1 & 2D & Logiq E10     \\
		
    \bottomrule
    \end{tabular}}%
\hfill
\hfill
\hfill
\renewcommand{\arraystretch}{1.5}
    \resizebox{0.48\columnwidth}{!}{
    \begin{tabular}{|c|c|c|c|c|c|c|}		
    \multicolumn{2}{l}{} & \multicolumn{2}{l}{\textbf{Test Data-set}} & \multicolumn{2}{l}{}\\		
		\toprule
    \textbf{Num} & \textbf{Anatomy} & \textbf{\#Images} & \textbf{Objects} & \textbf{Probe} & \textbf{Device} \\    
		\midrule
    1     & \makecell{{Cardiac} \\ {Short axis}}  & 2250 & TV annulus & 2D & VIVID E9/E95     \\
		\midrule
    2     & Fetal Head  & 4786 & HC & 2D & Voluson E8/E10  \\
		\midrule
    3     & Liver  & 95 & Liver & 2D & Logiq e     \\
		\midrule
    4     & Breast Lesions & 648 & Breast lesions & 3D & Public \cite{BUSI}     \\
		\midrule
    5     & \makecell{Muscolo-skeletal}  & 8000 & \makecell{Muscle tissue} & 2D & Public \cite{transverse_msk}    \\
		\midrule
    6     & \makecell{{Adult cardiac 4ch}} & 10000 & LV, LA & 2D & Public  \cite{CAMUS}   \\
		\midrule
		7     & \makecell{{Adult cardiac 2ch}} & 10000 & LV, LA & 2D & Public  \cite{CAMUS}   \\
		\bottomrule
    \end{tabular}%
}
\vspace{0.1 in}
\caption{ Left: Training data-sets; Right: Test data-sets}

\end{table*}%
\begin{figure*}[!b]
\centering
\includegraphics[width=\textwidth]{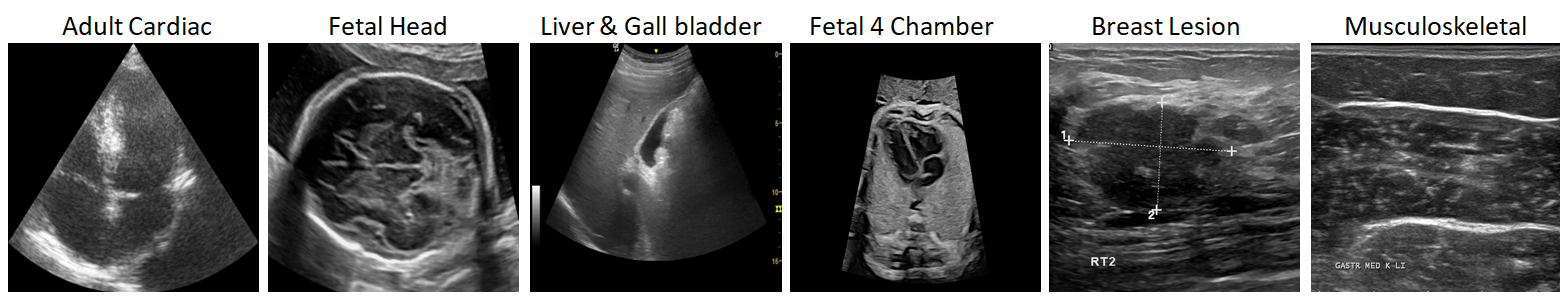}
\caption{Collection of images depicting variability of objects present in ultrasound.}
\label{fig:Ultrasound views}
\end{figure*}
\section{Training Data-sets}
\label{sec:Datasets}
\vspace{-0.1in}
Ultrasound imaging poses challenges to any AI model development, specifically 
for foundation models due to poor SNR and CNR compared to other imaging modalities. Moreover, the concept of ``objectness" do not translate well from natural images to ultrasound. As shown on few exemplar images in Fig. 3, the object boundaries are not often well defined, texture is highly varying and overall image quality has high inter-operator variability. To build generalized foundation models, training data-set has to be curated carefully to ensure sufficient diversity.  We primarily utilized data from a wide variety of commercially available devices for training. Our training data-set  (Table. 1) includes almost 200k images with 1) varying echogenicity: Hyper-echogenic structures like fetal cranium, hypo-echogenic regions, anechoic and fluid filled objects like fetal cardiac, 2) varying interfaces: objects with soft tissue boundaries like kidney, bone-tissue interface in fetal thorax 3) varying texture: homogenous texture like liver, thyroid, and heterogeneous texture in uterus. 4) varying image quality: 2-D( convex/ sector/ linear) probes, 3-D mechanical probes and electronic 4-D probes.  We demonstrate the value of curating such diverse data in results section.

\vspace{-0.2in}
\section{Experiments and Results}
\vspace{-0.1in}
We evaluate and report results of SonoSAM on 6 test datasets listed in Table 1. Note that these data-sets were chosen to test generalization capabilitiy of SonoSAM covering unseen anatomy (Adult Heart, Fetal Head, etc), unseen pathologies (Breast Lesions, MSK pathologies), different scanners (Logiq e), public data-sets (from different challenges). We compare the performance against four state-of-the-art methods namely RITM \cite{RITM}, FocalClick (\cite{FocalClick}), SAM \cite{SAM} and MedSAM \cite{medsam}. To automatically evaluate performance of these models with increasing number of user interactions, we start with centroid of the ground truth masks and add positive negative clicks depending on evolution of predictions. We note that MedSAM \cite{medsam} is not trained to work with clicks and hence we start with bounding box prompt. We use Dice Similarity Coefficient (DSC) between predicted and ground truth masks as the performance metric of choice. 
We report on the following set of derived metrics to analyze performance.
\begin{itemize} 
\item {Distribution of DSC scores averaged across images in each data-set versus increasing number of clicks from $1$ to $10$.}
\end{itemize}
\subsection{Results on 2D images}

\begin{figure*}[b!]
\includegraphics[width=1\textwidth]{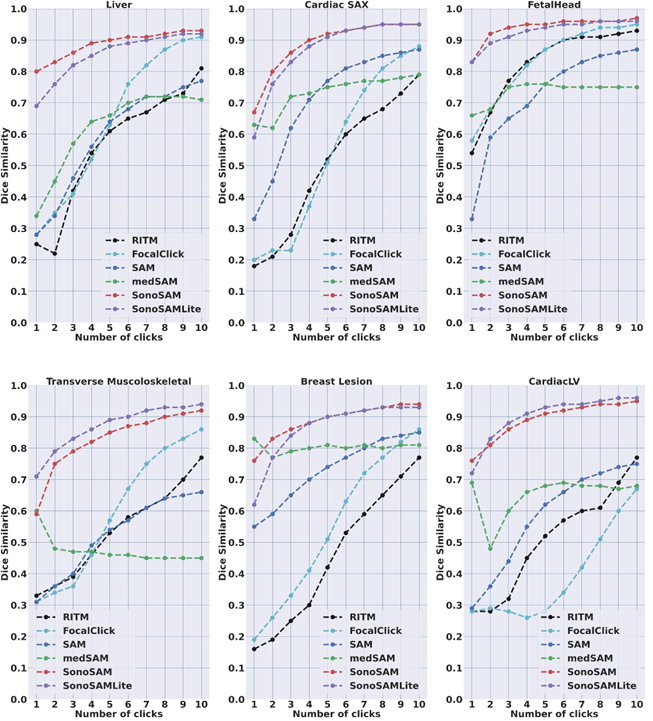}
\vspace{0.02in}
\caption{Figure showing Average DSC value for SonoSAM and SonoSAMLite models along with SOTA methods for increasing number of clicks on the 6 test data-sets. } 
\label{fig:ClicksVersusDice}
\end{figure*}

\subsubsection{a) SonoSAM achieves state-of-the-art performance on all test datasets:}
SonoSAM achieves $>90\%$ DSC on all data-sets and comfortably surpasses competing methods by a huge margin which struggle to cross even $80\%$ DSC. As shown in Table. 2, SAM model trained on natural images, under-performs significantly on ultrasound images often being poorer than SonoSAM in range of $8-41\%$ MaxDSC. Surprisingly, MedSAM which has been trained partly  on ultrasound images is often the worst performer amongst all models, despite 3 of these datasets being `in-domain' data-sets for MedSAM. Lack of training with clicks, severely hampers and infact deteriorates MedSAM's performance. FocalClick model, performs reasonably on two data-sets - Liver and Fetal Head but takes several clicks to get to meaningful results, as shown Figure ~\ref{fig:ClicksVersusDice}

\subsubsection{b) SonoSAM behaves predictably with user interaction}
\begin{figure*}[]
\label{fig:evolve}
\includegraphics[width=1\textwidth]{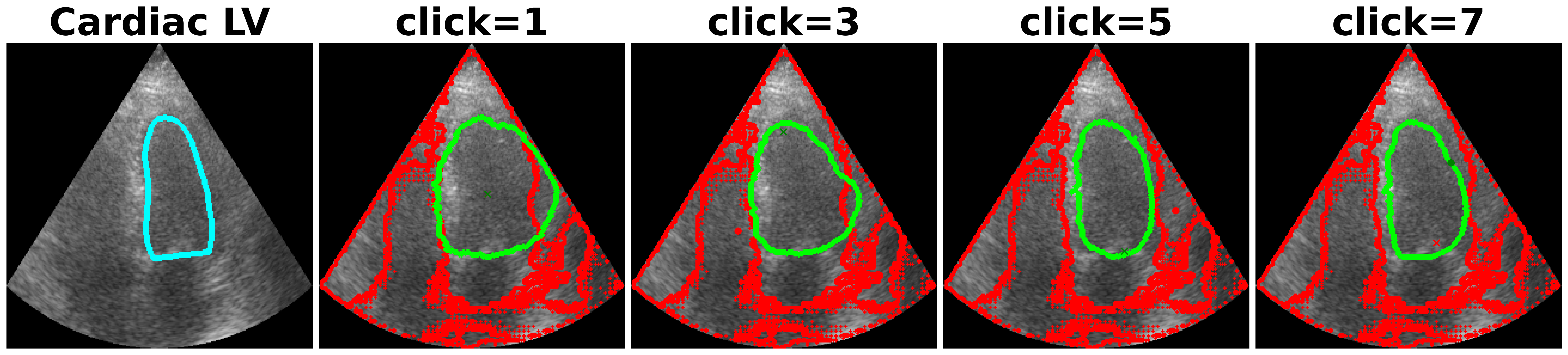}
\includegraphics[width=1\textwidth]{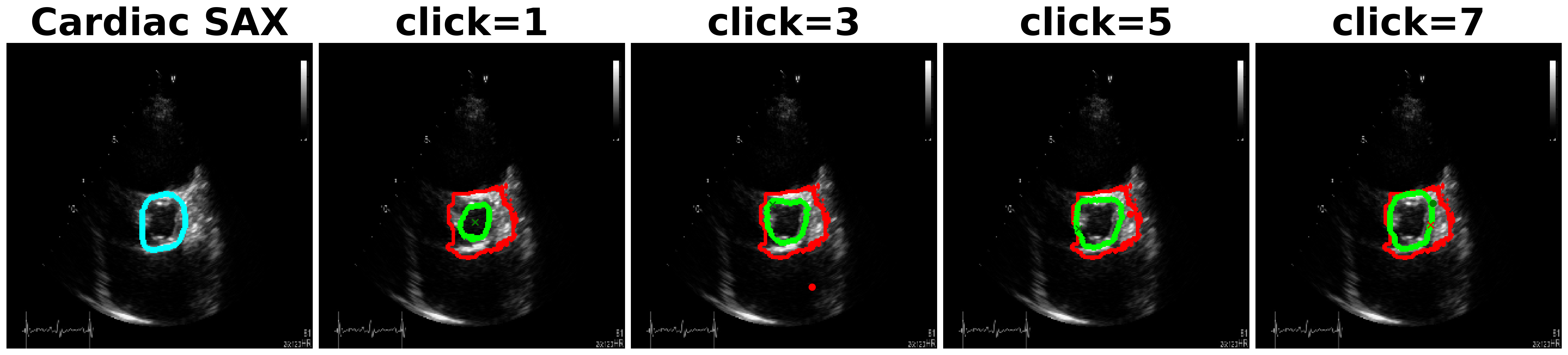}
\includegraphics[width=1\textwidth]{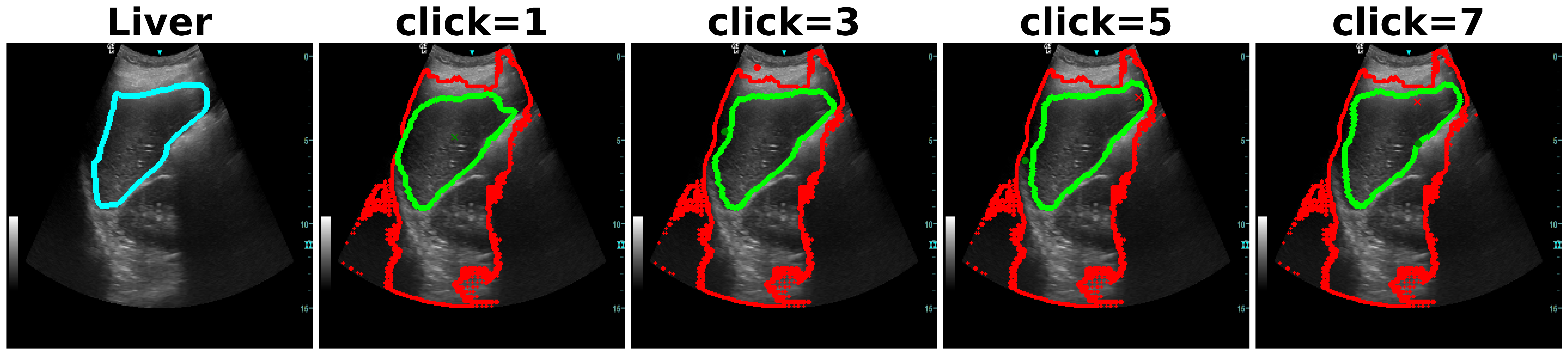}
\includegraphics[width=1\textwidth]{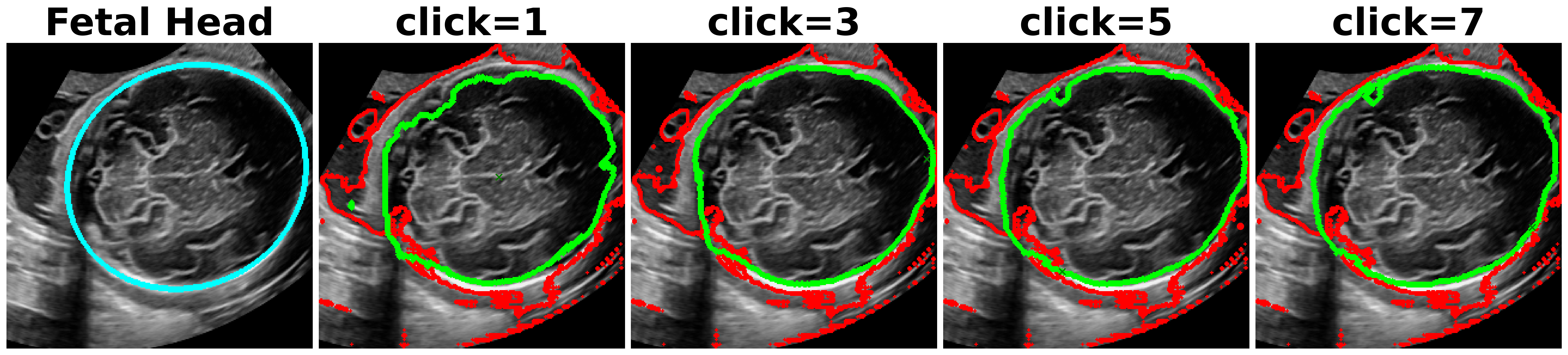}
\includegraphics[width=1\textwidth]{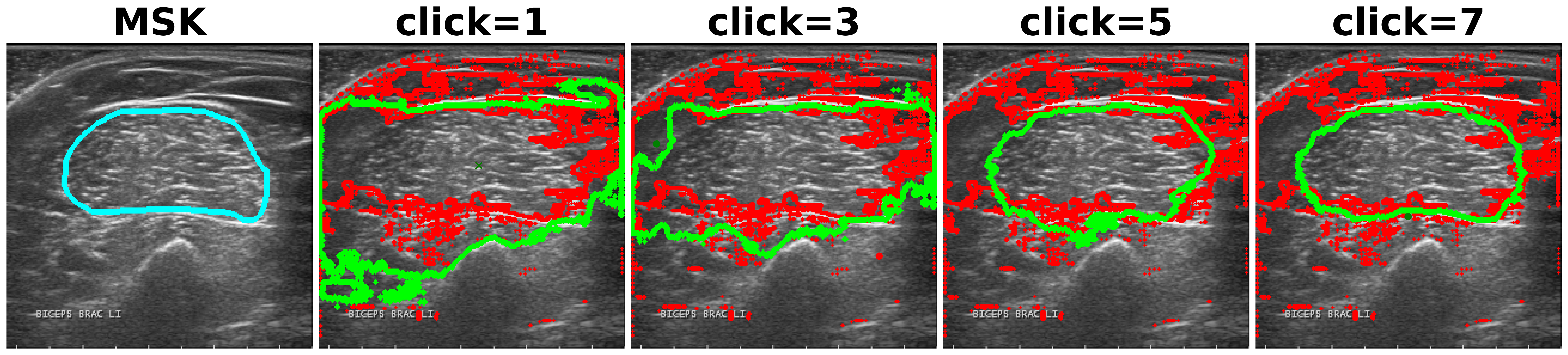}

\caption{Evolution of segmentation predictions for clicks - \textit{1,3,5,7} on 5 anatomies from test data-set. Legend - Red: SAM prediction, Green: SonoSAM, Cyan: GT}
\label{fig:resultImages}
\end{figure*}

One of the complaints with SAM model is that, the mask generated with user clicks (positive or negative clicks) are unintuitive and unpredictable. As shown in Figure ~\ref{fig:resultImages}, SAM often picks the entire FOV as the object and is unresponsive to multiple clicks. In contrast, SonoSAM's responses are predictable, as demonstrated by smooth progression of predicted contours as shown in Figure ~\ref{fig:resultImages}
\subsection{\textbf{Performance comparison of SonoSAM vs SonoSAM{\textbf{Lite}} }}
 \vspace{-0.1in}
As shown in Fig. \ref{fig:ClicksVersusDice} , SonoSAM{{Lite}} model performs very close to SonoSAM model on all of the 6 anatomical datasets for 1-10 clicks. For specific applications, ex. MSK , it is quite encouraging to note that SonoSAM{{Lite}} model outperforms the SonoSAM model. For a few anatomies ex. Fetal Head and Liver, the SonoSAM{{Lite}} model lags slightly in the initial few clicks, however the maximum DSC at the end of 10 clicks is very similar. 
 
\vspace{-0.2in}
\subsection{Evaluation on 2-D+t use-cases}

We study the utility of the approach to segment and propogate Left Ventricle and Left Atrium on 2-chamber and 4-chamber cardiac cine acquisitions with average number of frames varying around $20$.  On the first frame, SonoSAM is invoked to get contours of structures of interest. On subsequent frames, the deAoT tracking algorithm is called and results are monitored. Whenever, the tracked contour's dice overlap with ground truth is less than $90\%$, intervention with SonoSAM is performed to provide corrective clicks until satsifactory contours are again achieved. The process is repeated until the end of the frames. As mentioned earlier, metrics defined earlier that count number of interventions, quality of tracking at interventions and overall average number of clicks are reported. Table. \ref{table:track} reports quantitative performance on these metrics averaged across 30 subjects. The key take-aways are as follows:

\begin{itemize}
\item{The average number of interventions per loop is $<2$, which means that proposed tracking framework produces acceptable contours on all frames but 2 in the cine-loops.}Ventricle
\item{The average drop in dice before interventions is $<2\%$, which means that even on frames where dice overlap does not exceed $90\%$, the drop is not significant with only $2\%$ poorer.}
\item{The average number of clicks per loop is $\sim 1$, which is hugely significant. Roughly on a loop with 20 frames, user is expected to click only 20 times, proving the utility of the proposed approach.}
\end{itemize}


\begin{figure*}
\centering
\includegraphics[width = 0.7\textwidth]{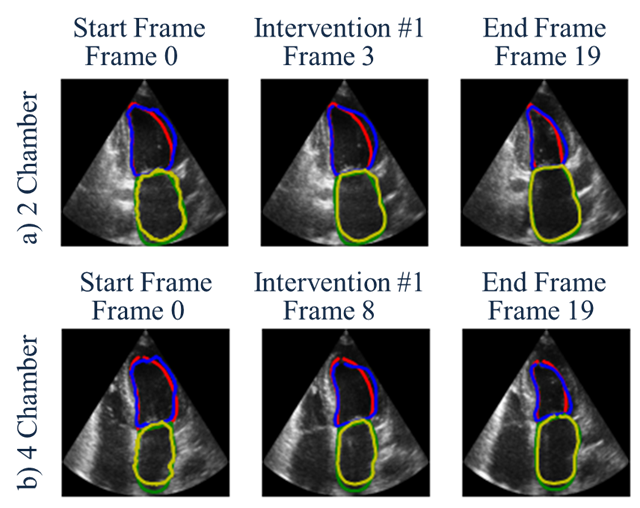}
\caption {Tracking results on Patient 7 with SonoSAM and deAoT \cite{track} a) On 2-chamber u/s cine-loop b) On 4-chamber u/s cine-loop for structures Left Atrium and Left Ventricle. Apart from start and end frames, the frame at which intervention with SonoSAM was required where tracking performance decreased below 90\% is shown. Legend: 1) Ground-truth a) Red: Left Ventricle b) Left Atrium 2) Prediction a) Blue: Left Ventricle b) Yellow: Left Atrium.  }
\label{fig:track1}
\end{figure*}

\vspace{-0.1in}
\begin{figure*}
\centering
\includegraphics[width = 1\textwidth]{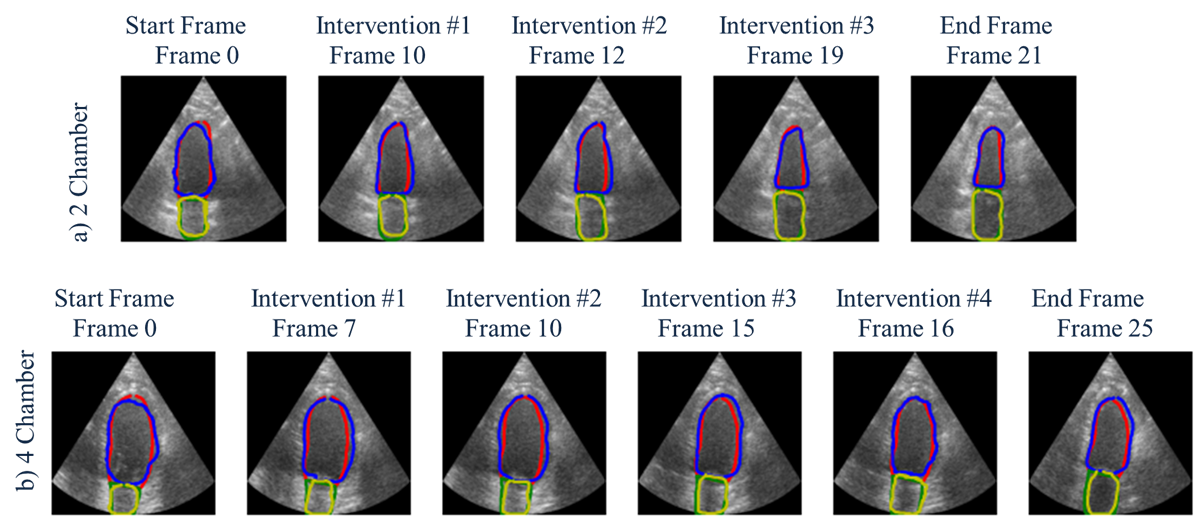}
\caption {Tracking results on Patient 5 with SonoSAM and deAoT \cite{track} a) On 2-chamber u/s cine-loop b) On 4-chamber u/s cine-loop for structures Left Atrium and Left Ventricle. Apart from start and end frames, the frames at which intervention with SonoSAM was required where tracking performance decreased below 90\% is shown. Legend: 1) Ground-truth a) Red: Left Ventricle b) Left Atrium 2) Prediction a) Blue: Left Ventricle b) Yellow: Left Atrium.  }
\label{fig:track2}
\end{figure*}

%
\vspace{-0.1in}
\begin{table*}[]
\centering
    \resizebox{0.50\textwidth}{!} {
\begin{tabular}{|c|c|c||c|c|}		
    \toprule
     \multirow{3}*{\makecell{{Anatomy/} \\ {Metric}}}& \multicolumn{2}{r|}{Adult 2 Chamber}  & \multicolumn{2}{|r|}{Adult 4 Chamber}  \\      
    \cmidrule{2-5}
    & \makecell{{Left} \\ {Ventricle}}  & \makecell{{Left} \\ {Atrium}}  & \makecell{{Left} \\ {Ventricle}}  & \makecell{{Left} \\ {Atrium}}   \\
		\midrule
		\makecell{{Avg. num of } \\ {interventions}}& 1.7	& 1.2 &	1.9	& 1.4\\				
		\midrule
		\makecell{{Avg. drop of DSC } \\ { before interventions}}& 1.80\%	& 1.30\% &	1.70\% &	1.80\%\\
		\midrule
		\makecell{{Avg. num of } \\ {clicks per loop}}& 1.1	& 0.9 &	1.0	& 1.0\\

    \bottomrule
    \end{tabular}}%
		\caption{Quantitative results on 3 tracking metrics. Results are averaged across 30 subjects. On average, combination of SonoSAM + deAoT requires roughly 1 click per loop to get $90\%$ or more dice overlap on all frames in the cine-loops. }
		\label{table:track}
\end{table*}

Qualitative results are shown in Fig. \ref{fig:track1} and \ref{fig:track2}. These figures show start and end frames of the cine-loops for both 2 chamber and 4 chamber views. Additionally, the frames on which interventions were required (i.e.)  when tracked contour's dice falls below $90\%$ is shown. In Fig. \ref{fig:track1}, the combination of sonoSAM and tracking was sufficient in all frames except: Frame 3. Fig. \ref{fig:track2} depicts a tougher case, where number of interventions were three and four on 2 chamber and 4 chamber loops respectively. The averaged statistics across 30 patients are shared in Table 3. 
\section{Conclusion}
\vspace{-0.1in}

While large vision models like SAM \cite{SAM} have demonstrated extraordinary results in natural images, there have been very little models in the medical imaging world.In this work, we present SonoSAM and SonoSAMTrack - a  foundational model for ultrasound which for the first time - achieves performance on ultrasound images on-par with vision models in  natural images. We demonstrate that by carefully curating training data-set of sufficient diversity, images in order of $200k$ is sufficient to get SOTA results. We illustrate that partial fine-tuning of large vision models - building domain-specific decoder is tractable and promising solution for ultrasound images. While success in computer vision has been achieved with humungous models, we show that with knowledge distillation, SonoSAMLite model of meager $30M$ parameters can perform as good as, if not outperform large models. We also demonstrate SonoSAMTrack, that enables 2D+t dense segmentations on top of SonoSAM. In our future research, we plan to analyze failure modes in detail, explore full fine-tuning, anatomy-specific smaller models, enabling bounding box and text prompts.
 
%
%
%

\end{document}